\documentclass{ws-procs975x65}
\usepackage{wrapfig}
\def\beq{\begin{equation}}
\def\eeq{\end{equation}}
\newcommand{\eqb}{\begin{eqnarray}}
\newcommand{\eqe}{\end{eqnarray}}
\newcommand{\ns}{\nu_{\rm s}}
\newcommand{\ep}{\epsilon_{\nu, p}}

\newcommand{\tcr}{t_{\rm cr}}

\newcommand{\tpesc}{t_{\rm p,esc}}
\newcommand{\teesc}{t_{\rm e,esc}}

\newcommand{\doppler}{\delta}
\newcommand{\en}{\epsilon_{\nu}}
\begin{document}
\title{The many faces of blazar emission in the context of hadronic models}
\author{Maria Petropoulou$^1$, Stavros Dimitrakoudis$^2$, Paolo Padovani$^3$, Elisa Resconi$^4$, Paolo Giommi$^5$ and Apostolos Mastichiadis$^6$}
\address{$^1$Department of Physics and Astronomy, Purdue University,
West Lafayette, IN 47907, USA\\
% $^*$E-mail: mpetropo@purdue.edu\\
$^2$Department of Physics, University of Alberta, Edmonton,
Alberta, Canada\\
$^3$European Southern Observatory, D-85748 Garching bei M{\"u}nchen, Germany\\
$^4$ Technische Universit{\"a}t M{\"u}nchen, D-85748  Garching bei M{\"u}nchen, Germany\\
$^5$ ASI Science Data Center, via del Politecnico s.n.c., I-00133 Roma Italy \\
$^6$ Department of Physics, University of Athens, 15783 Zografos, Greece}
\begin{abstract}
We present two ways of modeling the spectral energy distribution of blazars in the hadronic context and discuss the predictions of each ``hadronic variant'' on the spectral shape, the multi-wavelength variability, the cosmic-ray flux, and the high-energy neutrino emission. Focusing on the latter, we then present an application of the hadronic model to individual BL Lacs that were recently suggested to be the counterparts of some of the IceCube neutrinos, and conclude by discussing the contribution of the whole BL Lac class to the observed neutrino background.
\end{abstract}


\bodymatter

%%%%%%%%%%%%%%%%% now a standard article style for the most part
% \vspace{-0.1cm}
\section{Introduction}
Blazar jets have long been considered as candidate sites of cosmic-ray acceleration to the
highest energies observed ($\sim5\times 10^{20}$ eV). In the light of the recent IceCube neutrino detections, the hadronic model for blazar emission becomes more relevant than ever before. We compare the predictions of two variants of hadronic models for the the blazar spectral energy distribution (SED) by using the nearby BL Lac Mrk 421 as our testbed.  
\section{The Model}
\label{sec:model}
We adopt a one-zone leptohadronic model for the blazar emission, where 
the low-energy emission of the blazar SED 
is attributed to synchrotron radiation of relativistic electrons and 
the observed high-energy (GeV-TeV) emission 
is assumed to have a photohadronic origin.

We assume that the region responsible for the blazar emission can be described 
as a spherical blob of radius $R$, containing a 
tangled magnetic field of strength $B$ and moving towards us 
with a Doppler factor $\doppler$. Protons and (primary) electrons are assumed to be accelerated into power-law energy distributions and to be 
subsequently injected  isotropically in the volume of the blob with a constant rate. All particles are assumed to escape from the emitting region in a
characteristic timescale, which is set equal to the photon crossing time of the source, i.e. $\tpesc=\teesc=R/c$. 

Photons, neutrons and neutrinos ($\nu_{\mu}, \nu_e$) complete the set of the five stable populations, 
that are at work in the blazar emitting region. Pions ($\pi^{\pm}, \pi^0$), muons ($\mu^{\pm}$) and kaons ($K^{\pm}, K^0$)
constitute the unstable particle populations, since 
they decay into lighter particles.  
The production of pions is a natural outcome of photohadronic interactions between
the relativistic protons and the internal photons; the latter are predominantly synchrotron photons
emitted by the primary electrons. 
\begin{wrapfigure}{r}{0.5\textwidth}
\centering
\vspace{-0.1in}
\includegraphics[width=0.49\textwidth,height=0.3\textwidth]{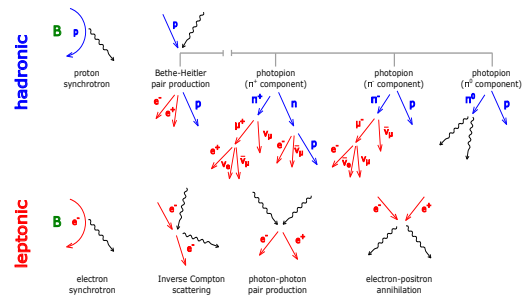}
% \vspace{-0.1in}
  \caption{Schematic illustration of the main hadronic and leptonic processes that are included
  in our numerical treatment.}%
  \label{fig:processes}
\end{wrapfigure}
The decay of $\pi^{\pm}$ results in the injection of secondary relativistic 
electron-positron pairs ($\pi^{+}\rightarrow \mu^{+}+\nu_{\mu}$, $\mu^{+}\rightarrow e^{+}+\bar{\nu}_{\mu}+\nu_{\rm e}$),
whose synchrotron emission emerges in the GeV-TeV regime, for a certain range of parameter values.
$\pi^0$ decay into very high energy (VHE) $\gamma$-rays (e.g. $E_{\gamma}\sim10$~PeV, for a parent proton with energy
$E_{\rm p}=100$~PeV), and those are, in turn, susceptible to photon-photon ($\gamma \gamma$) absorption and can 
initiate an electromagnetic cascade\cite{mannheim91}. As the synchrotron self-Compton emission from primary electrons
may also emerge in the GeV-TeV energy band, the observed $\gamma$-ray emission can be totally or partially explained by photohadronic processes, depending on the specifics of individual sources\cite{petroetal15}. 

The interplay of the processes (see Fig.~\ref{fig:processes}) governing the evolution 
of the energy distributions of the five stable particle populations is 
formulated with a set of five time-dependent, energy-conserving kinetic equations. 
To simultaneously solve the coupled kinetic equations for all 
particle types we use the time-dependent code described in Ref.~\refcite{DMPR2012}. 

\section{Hadronic Modeling Of The BL Lac Mrk 421}
Mrk 421 is one of the nearest ($z=0.031$) and
brightest BL Lac sources in the VHE ($E_{\gamma}>200$~GeV) sky and extragalactic X-ray sky, which makes it an
ideal target of multi-wavelength observing campaigns. Using Mrk 421 as our testbed, we present two ways of modeling
the blazar SED in the hadronic context, namely the LH$\pi$ and LHs models. \tref{table1} summarizes the main features of those models, while details about the spectral shape and variability, the neutrino and cosmic-ray emission are presented in the following paragraphs.
\begin{table}
\tbl{Comparison of the hadronic model variants LH$\pi$ and LHs.}
{\begin{tabular}{@{}c c c@{}}
\toprule
                        & LH$\pi$ model & LHs model \\
\colrule
UV-to-X-rays      & primary $e^{-}$ synchrotron & primary $e^{-}$ synchrotron  \\
GeV-to-TeV $\gamma$-rays     & secondary $e^{-}$ synchrotron & $p$ synchrotron \\
Dominant energy density & proton  &   magnetic \\
Jet power (erg/s) & $\sim 10^{48}$ & $\sim 10^{46}$ \\
Maximum proton energy  &  $\sim 20$ PeV    & $\sim 20$ EeV \\
Maximum neutrino energy &  $\sim 1$ PeV   &  $\sim 1$ EeV \\
X-ray flux vs. TeV flux & quadratic & linear\\ 
\botrule
\end{tabular}
}
\label{table1}
\end{table} 
% \vspace{-0.3in}
\subsection{Photon, neutrino and cosmic-ray spectra}
\begin{figure*}
\centering
\includegraphics[width=0.31\textwidth]{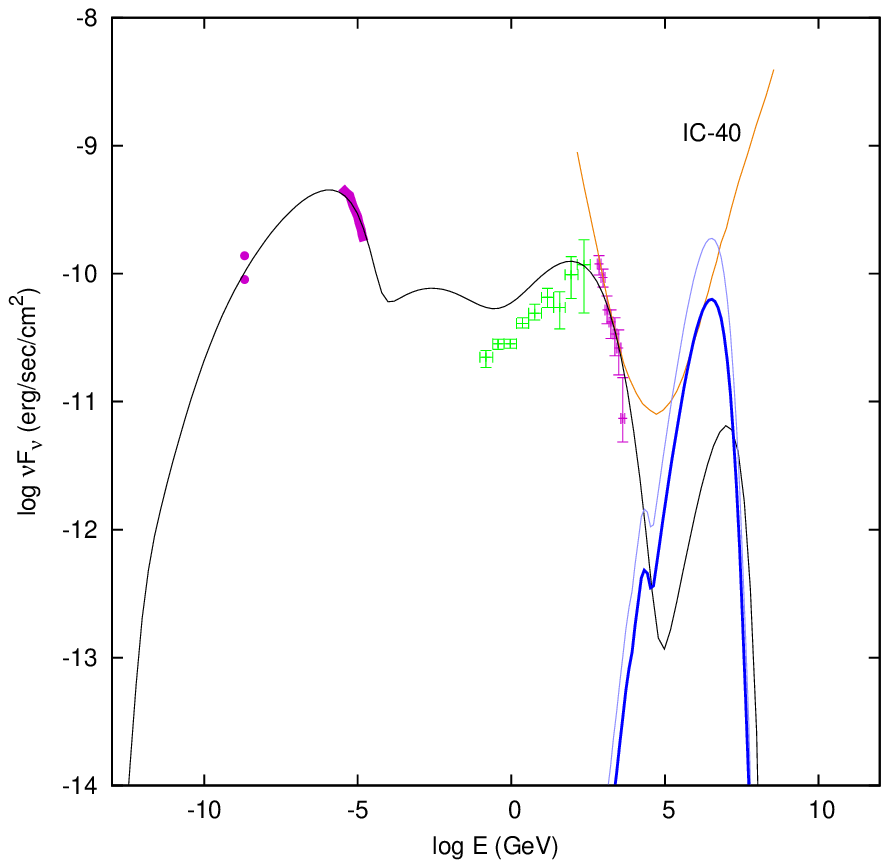}
%   \hspace*{0.1in}
\includegraphics[width=0.31\textwidth]{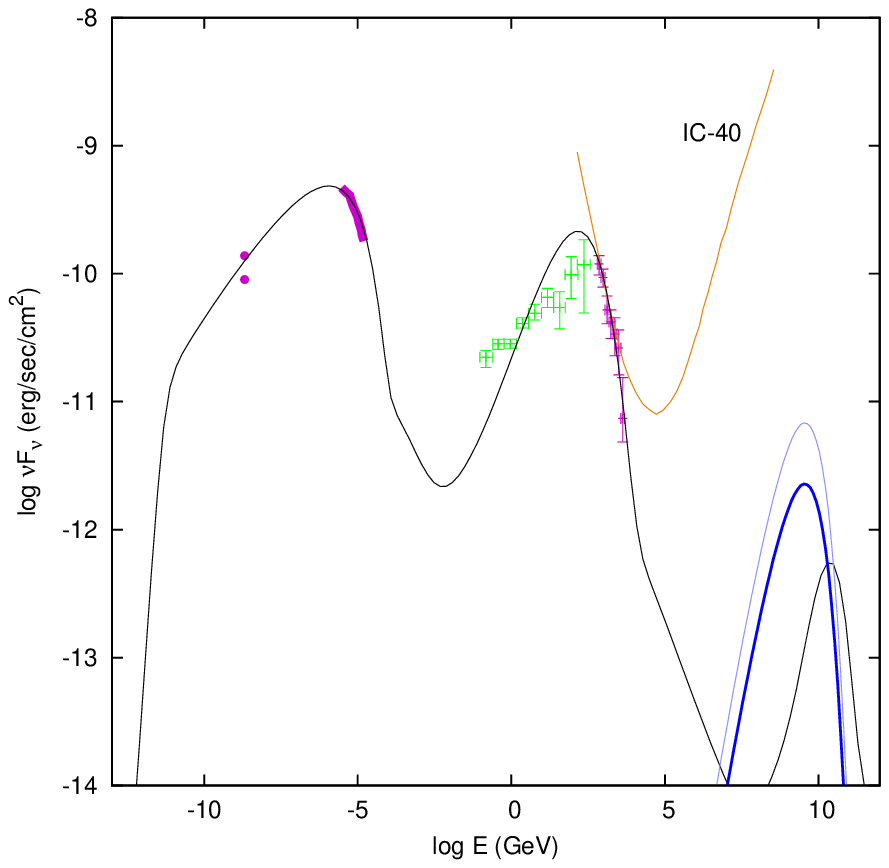}
%   \hspace*{0.1in}
\includegraphics[width=0.31\textwidth]{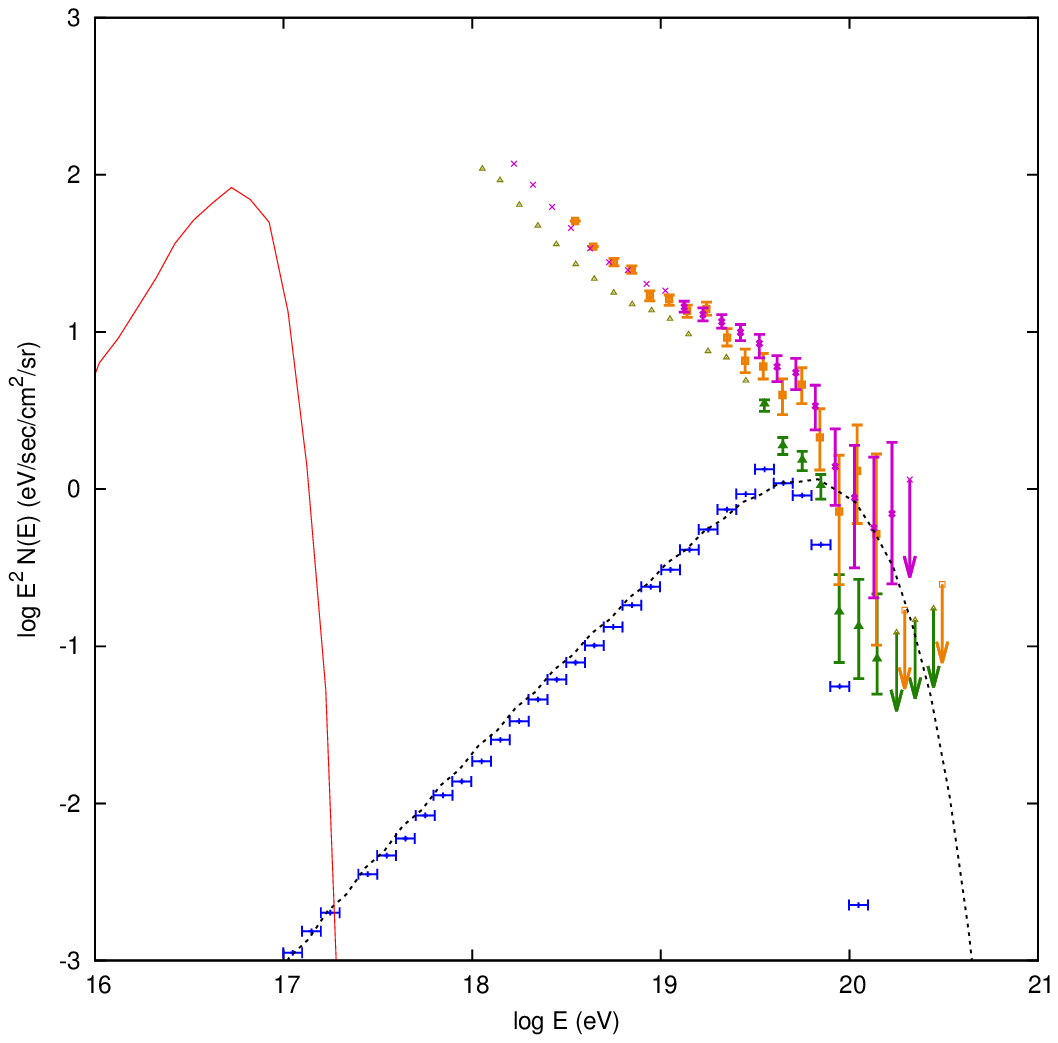}
  \caption{{\sl Left and middle panels}: spectra of photons (black line) fitting the March 22nd/23rd 2001
observation of Mrk 421 (purple points), neutrinos of all flavors (grey line) and $\nu_\mu+\bar{\nu}_\mu$ flux
(thick blue line) according to the LH$\pi$ and the LHs models respectively.
Fermi observations (green points) are not simultaneous with the rest of the data and thus not included in the fit. 
The 40-String IceCube limit\cite{tchernin13} for $\nu_\mu$ is plotted with an orange line. {\sl Right panel}: cosmic-ray (proton) spectra resulting from neutron decay and obtained within the LH$\pi$ (red line) and LHs (blue dashed line) models. For the latter, the cosmic-ray
spectra obtained after taking into account propagation effects using the numerical code CRPropa
2.0\cite{kampert13} are also shown (blue crosses). Different symbols are used for the cosmic-ray energy flux measurements by Auger,
HiRes-I, and Telescope Array.}%
  \label{fig:spectra}
% \end{center}
\end{figure*}

In the LH$\pi$ model (Fig.~\ref{fig:spectra}, left panel) the proton synchrotron emission is suppressed, whereas the photopair and photopion components are prominent. This is the result of a low magnetic field in combination with a high proton luminosity. The SED does not have the usual double-humped
appearance as synchrotron photons from the photopair secondaries produce a broad hump at MeV energies (see also Ref.~\refcite{petromast15}). The energetic requirements of this model are high (see \tref{table1}), while most of the energy is carried by the highest energy particles.
Although the radiative efficiency of the model is low ($\sim 10^{-5}$), the high proton luminosity leads to a substantial neutrino flux that
is of the same order as the TeV γ−rays. Interestingly, the expected $\nu_\mu+\bar{\nu}_\mu$ flux, which peaks at $\sim 3.3$~PeV, is just
under the sensitivity of the IC-40 detector (orange line). The cosmic-ray proton spectrum resulting from neutron decay peaks at $70$~PeV (Fig.~\ref{fig:spectra}, right panel).  This is just an upper limit of  what it would appear at Earth since we have not taken into account CR diffusion,
which is important for energies $<10^{17}$ eV. At any rate, our values, even as an upper
limit, are well below the observed CR flux at such energies.
In the LHs model, the high magnetic field coupled with a low proton injection luminosity leads to 
a suppressed photohadronic component. The SED has two well-defined peaks,
both from synchrotron radiation of electrons and protons at UV/X-rays and GeV/TeV $\gamma$-ray energies, respectively. 
This also results in a low neutrino flux (a factor of 10 less than the TeV $\gamma$-ray flux). The peak of the neutrino flux emerges at energies of $\sim 0.1$~EeV due to the high values of the magnetic field and of the maximum proton energy (see \tref{table1}).
The higher value of the maximum proton Lorentz factor used in the SED fitting
makes the discussion about ultra-high energy cosmic-ray (UHECR) emission more relevant. The propagation of
UHE protons in a uniform intergalactic $1$pG magnetic field and their energy losses
from interactions with the cosmic microwave and infrared–optical backgrounds were
modeled using CRPropa 2.0. The resulting spectra (blue crosses in right panel of
Fig.~\ref{fig:spectra}) peak at $\sim 60$ EeV and they are just below the present UHECR flux limits
in the energy range $30-60$ EeV.
\subsection{Variability}
\begin{figure}
\centering
\includegraphics[width=0.42\textwidth]{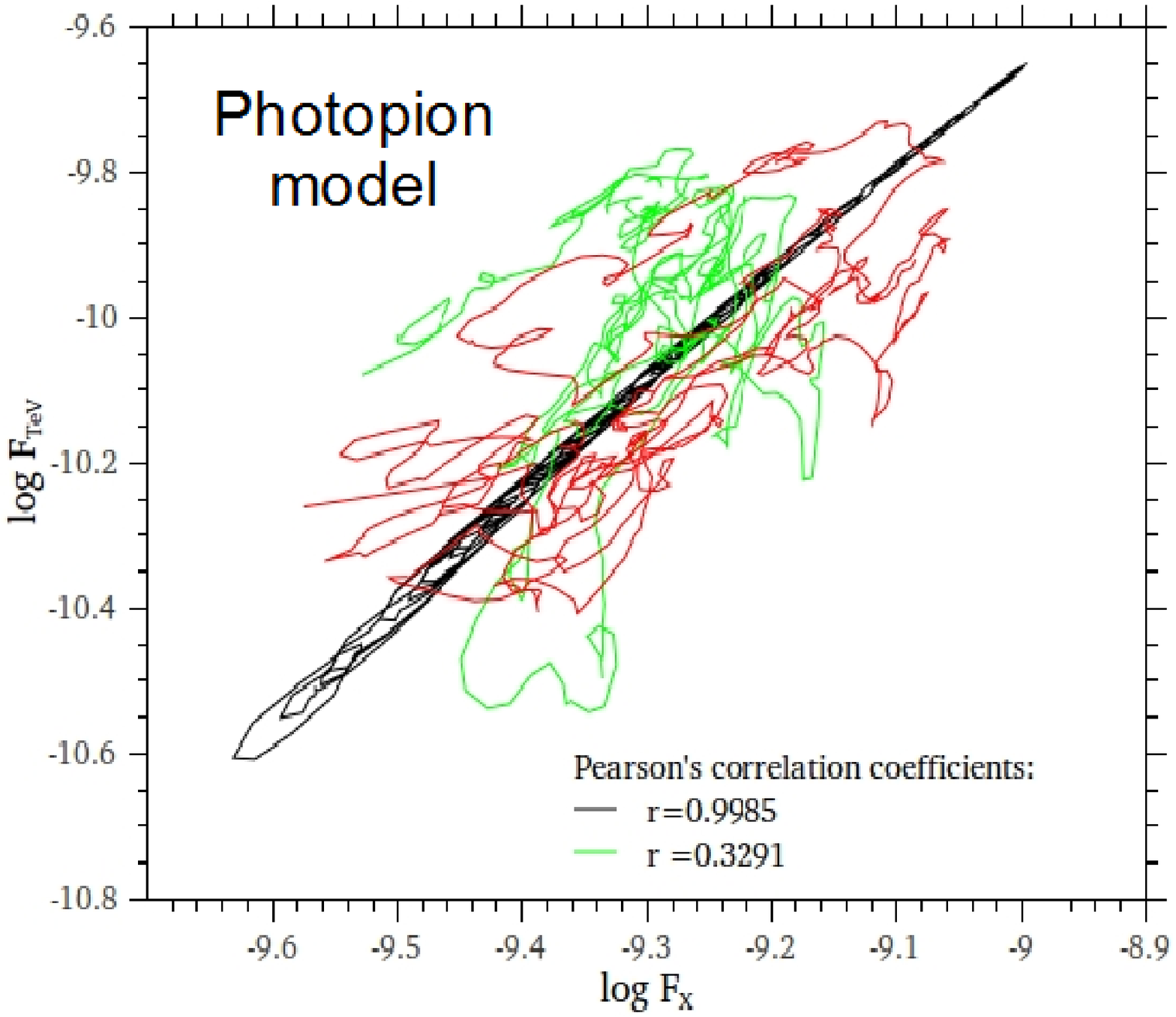}
\hspace{0.2in}
\includegraphics[width=0.45\textwidth]{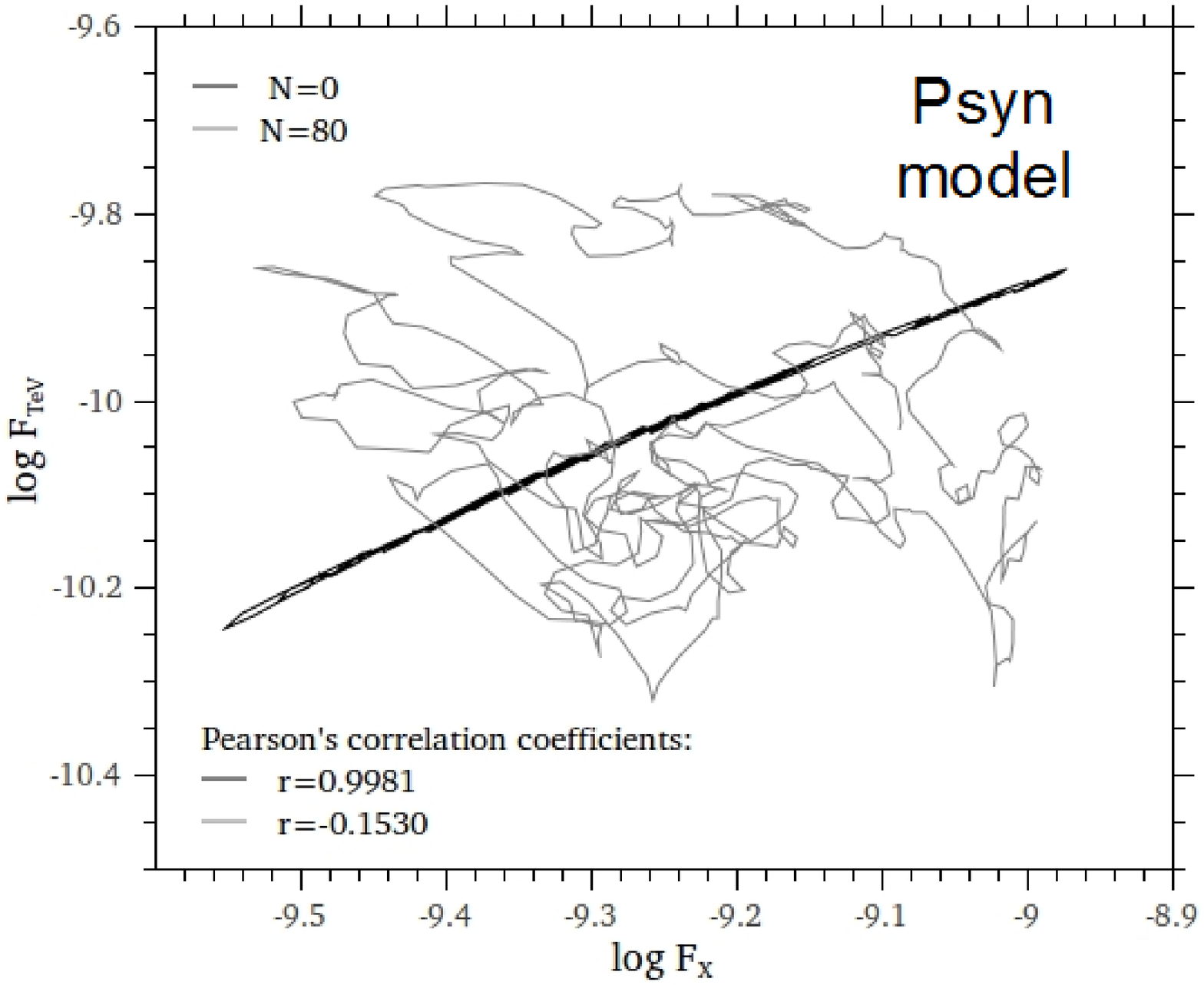}
%   \hspace*{0.1pt}
  \caption{TeV $\gamma$-ray flux vs. X-ray flux as obtained in the LH$\pi$ (left panel) and LHs (right panel) models by varying the injection luminosity of 
  primary electrons and protons. We considered the cases of uncorrelated variations (green line) as well as of correlated variations with no time-lag (black lines) and a positive time-lag of 80$\tcr$ (red and grey lines in the left and right panels, respectively). }%
  \label{fig:variability}
% \end{center}
\end{figure}

Recently, the variability signatures expected in the framework of hadronic models have been studied in Ref.~\refcite{mastetal13} by introducing small-amplitude variations to one (or more) model parameters around their time-averaged values. In particular, the temporal variations in the fitting parameter $y$ were modeled as  random-walk  changes of the form $y_{i} \equiv y(t_i) = y_0 \left(1+0.05\alpha_i \right)$, where $\alpha_{i+1}= \alpha_i + (-1)^{\kappa}$; here,  $\kappa$ is a uniformly distributed random integer number in the range (0,10). An indicative example is presented in Fig.~\ref{fig:variability}, where the varying model parameters are the proton and primary electron injection luminosities. A strong correlation between the X-ray and TeV $\gamma$-ray fluxes is found in the LH$\pi$ model. Moreover, a quadratic relation between the TeV and X-rays fluxes is found, similarly to the leptonic SSC model. In the LHs model, the correlation is present but not as strong as in the LH$\pi$ model, while the TeV 
$\gamma$-rays vary linearly with respect to X-rays. 

\section{Neutrino emission from individual BL Lacs}
In Ref.~\refcite{padovaniresconi14} the authors have recently searched for plausible astrophysical counterparts 
within the median error circles of IceCube neutrinos using a model-independent method and derived the most probable counterparts for 9 out of the 18 neutrino events of their sample. Interestingly, these include 8 BL~Lac objects (6 with measured redshifts), amongst which the nearest blazar, Mrk~421, and two pulsar wind nebulae.  
The (quasi)-simultaneous SEDs of those 6 BL Lacs, namely
Mrk~421, 1ES~1011+496, PG~1553+113, H~2356$-$309, 1H~1914$-$194, and 1RXS~J054357.3$-$553206, were fitted\cite{petroetal15} 
with the leptohadronic model described in Section~\ref{sec:model}. The all-flavor neutrino fluxes derived by the model
are presented in Fig.~\ref{fig:neutrinos}(a). 
\def\figsubcap#1{\par\noindent\centering\footnotesize(#1)}
\begin{figure}
 \centering
\begin{minipage}[l]{0.49\textwidth}
\includegraphics[width=1\linewidth,height=0.75\linewidth]{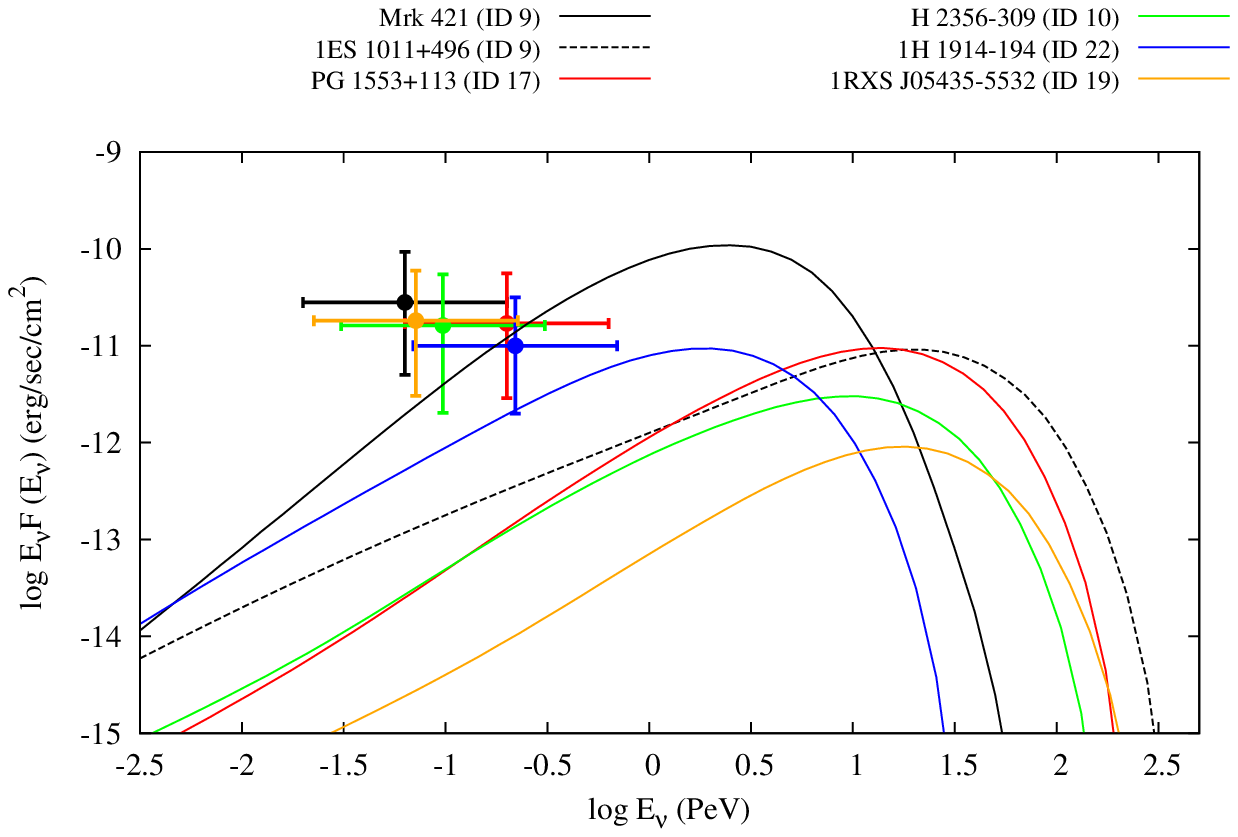}\figsubcap{a}
\end{minipage}
% \hspace*{0.1in}
\begin{minipage}[r]{0.49\textwidth}
\includegraphics[width=0.9\linewidth]{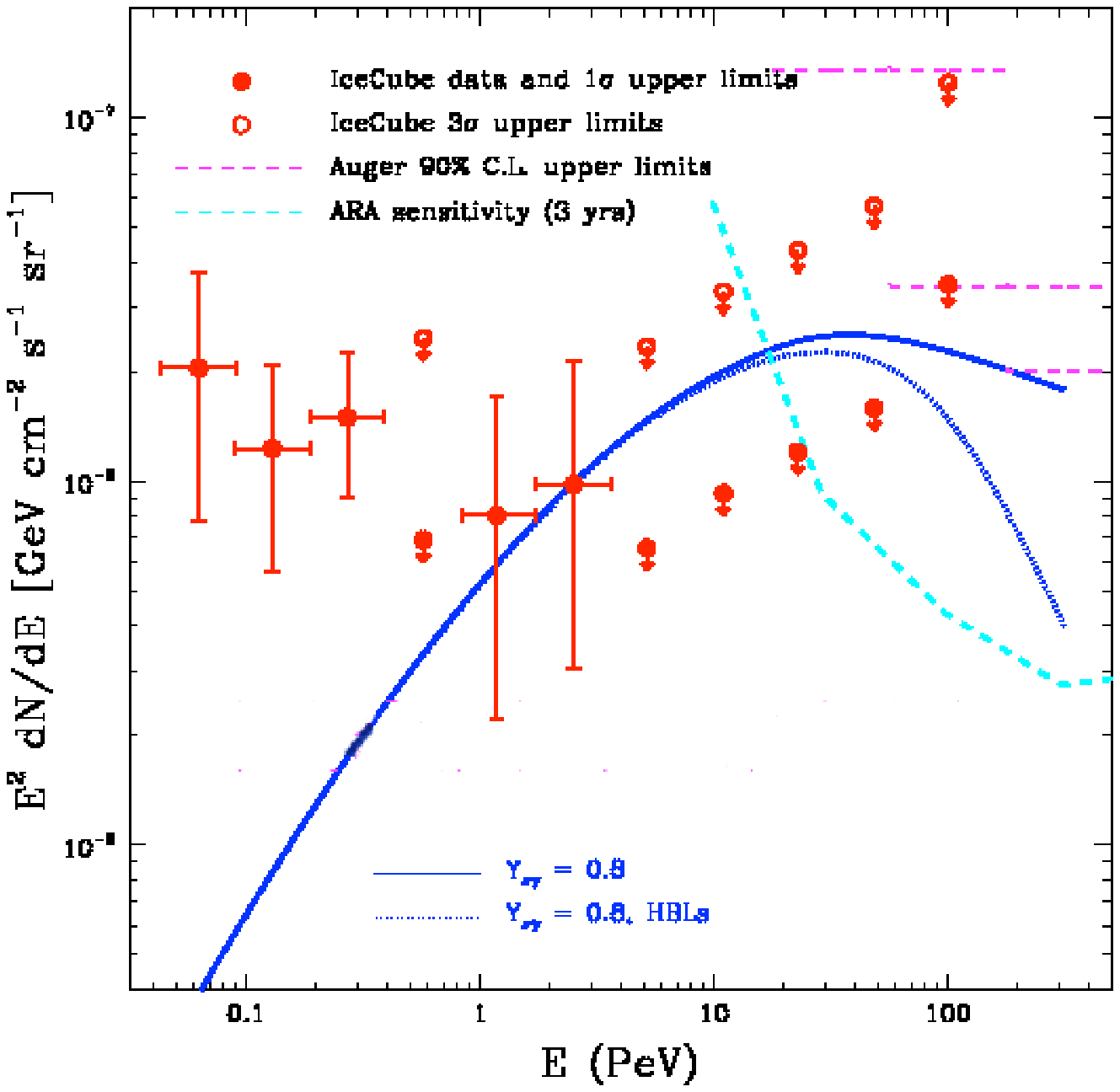}\figsubcap{b}
\end{minipage}
\caption{\footnotesize{(a) Comparison of the model (lines) and the observed (circles) neutrino fluxes as defined in Ref.~7 for the six BL~Lacs of the sample. The Poissonian 1$\sigma$ error bars for each event are also shown. (b) The predicted neutrino background (per neutrino flavor) from all BL  Lacs (blue solid line) and from high-frequency peaked BL Lacs  (HBL) only (blue dotted line)
  for $Y_{\nu\gamma} = 0.8$ and $E_{\rm break} = 200$ GeV, $\Delta \Gamma = 0.5$.
  The curves correspond to the mean value of ten different simulations. The (red) filled points are the data
  points from Ref.~\cite{aartsen14}, while the open points are the
  $3\sigma$ upper limits. The upper (magenta) short dashed line represents
  the 90\% C.L. upper limits from Auger\cite{Auger2013} while the lower
  (cyan) short dashed line is the expected three year sensitivity curve for
  the Askaryan Radio Array\cite{ARA12}.}}
 \label{fig:neutrinos}
\end{figure}
According to the model-independent analysis of Ref.~\refcite{padovaniresconi14}, neutrino event 9
has two  plausible astrophysical counterparts:  the BL Lacs Mrk~421 and 1ES~1011+496.  
The differences between the neutrino fluxes originate from the differences in their SEDs. In this regard,
the case of neutrino event 9 reveals in the best way how detailed information from 
the photon emission may be used to lift possible degeneracies between multiple astrophysical counterparts.
As the neutrino spectrum for 1ES~1011+496 (dashed line in Fig.~\ref{fig:neutrinos}(a)) is an upper limit, our results strongly favor Mrk~421 against 1ES~1011+496.

In all cases, the model-derived neutrino flux at the energy bin of the detected neutrino
 is  below the 1$\sigma$ error bars, but still within the 3$\sigma$ error bars. 
 Although the association of these sources cannot be,  strictly speaking, excluded at the present time,
 blazars Mrk~421 and 1H~1914-194 are the two most interesting cases, 
 because their association with the respective IceCube events
 can be either verified or disputed in the near future.  
Figure~\ref{fig:neutrinos}(a) demonstrates that the  model-derived neutrino spectra 
from blazars with different properties are similar in shape. We may thus model 
the observed differential neutrino plus anti-neutrino
($\nu+\bar{\nu}$) flux of all flavors ($F_{\nu}(\en)$) as $F_{\nu}(\en) = F_0 \en^{\beta} \exp\left(-\frac{\en}{E_0} \right)$,
where $\langle \beta \rangle \sim 0.34$ and $E_0$ is in good approximation equal to the peak energy of the neutrino spectrum, namely
$\ep(\delta, z, \ns) \simeq 17.5 \ {\rm PeV} (1+z)^{-2}  \left(\delta/10\right)^2 \left(10^{16}{\rm Hz}/\ns\right)$.
In the above, $\delta$ is the Doppler factor, $z$ is the source redshift and $\ns$ is the {\it observed} synchrotron
peak frequency. The luminosity from the photopion component is directly connected to that of $\sim 2 - 20$~PeV neutrinos. Thus, 
our approach allows us to associate the observed blazar $\gamma$-ray flux with the expected all-flavor neutrino flux as
$F_{\nu, \rm tot} = Y_{\nu \gamma} F_{\gamma}\left(> E_{\gamma} \right)$, where $E_{\gamma}=10$~GeV and $Y_{\nu \gamma}$ is a factor
that includes all the details about the efficiency of photopion interactions; for example,  
$Y_{\nu \gamma} \ll 1$ implies an SSC origin for the blazar $\gamma$-ray emission. The normalization $F_0$ can be then inferred from the above.
% Eqs.~(\ref{eq:eq0}) and (\ref{eq:eq2}).

\section{The neutrino background from BL Lacs}
The calculation of the neutrino background (NBG) from {\it all} BL Lacs 
requires detailed knowledge of the blazar population in terms of
$\ns$, $\delta$, $\gamma$-ray fluxes and redshift. All these parameters, and many more,
are available in the Monte Carlo simulations presented in a series of papers by Giommi \& Padovani (e.g. Refs.~\refcite{paper1,paper3}). We note that the blazar SEDs in the simulations are extrapolated to the VHE band by using the simulated {\sl Fermi} fluxes and spectral indices and assuming a break at $E =E_{\rm break}$ and a steepening of the photon spectrum by $\Delta \Gamma$ (for details, see Ref.~\refcite{paper3}).

We assign to each blazar in the simulation a  neutrino spectrum. Since $E_0$ is fully determined for a given set of $\delta, z, \ns$ and $\beta$ covers a narrow range, we are left only with $Y_{\nu \gamma}$ as a possible ``tunable'' parameter. 
Then, we compute the total NBG  as $\int_{F_{\min}}^{F_{\max}}
F~(dN/dF) dF$ where $dN/dF$ is the differential number counts
and $F_{\min},F_{\max}$ are the fluxes over which these extend. To obtain
the NBG per neutrino flavor we divide our results by three. Finally, we perform ten simulations and  
calculate their average in order to smooth out the ``noise'' inherent to the Monte Carlo
simulations. 

The predicted NBG from BL Lacs is presented in Fig.~\ref{fig:neutrinos}(b) for $Y_{\nu\gamma} = 0.8$, $E_{\rm break} = 200$ GeV, and $\Delta \Gamma = 0.5$. We find that BL Lacs as a class (blue solid line) can easily explain the whole NBG  at $E_{\nu} \gtrsim 0.5$ PeV,
while they do not contribute much ($\sim 10\%$) at  lower energies. 
At $E_{\nu} \lesssim 30$~PeV most of the contribution to the NBG comes from  high-frequency peaked BL Lacs (HBL) (blue dotted line).
Although HBL represent a small fraction ($\sim 5\%$) of the BL Lac population,
they dominate the neutrino output up to $\approx 30$ PeV due to their relatively high
$\gamma$-ray, and therefore neutrino, fluxes and powers. According to preliminary calculations 
our results up to $\sim 1-2$~PeV are not sensitive on whether $Y_{\nu\gamma}$ is constant
or dependent on the blazar $\gamma$-ray luminosity. However, assuming an anti-correlation 
between $Y_{\nu \gamma}$ and $L_{\gamma}(>10 \rm GeV)$, we find that the predicted NBG at $E_{\nu}\gtrsim 5$~PeV is in tension with 
the 3$\sigma$ IceCube upper limits and the 90\% C.L. upper limits from Ref.~\refcite{Auger2013}. Thus, this  hypothesis is ruled out.

The model prediction on the detectability of 2~PeV$< \en <10$ PeV neutrinos for the NBG shown in Fig.~\ref{fig:neutrinos}(b) 
is $N_{\nu} \sim 4.6$ without taking into account the Glashow resonance (and $N_{\nu} \sim 7$, otherwise).
This calculation is based on the effective areas from Ref.~\refcite{aartsen13}. Since the model NBG  peaks 
at $\en > 10$ PeV, we expect 2-3 additional events up to $\sim 100$ PeV after 
making an educated guess on the effective areas above 10 PeV. 
Given that the 3$\sigma$ upper limit for 0 events is 6.6\cite{gehrels86}, the prediction of $N_{\nu} \sim 6.6-7.6$ is 
close to being inconsistent with the IceCube non-detections.
However, $Y_{\nu\gamma} = 0.8$ is likely an upper limit. This was derived, in fact, from a small sample of BL
Lacs, which may represent the tip of the iceberg in terms of neutrino
emission, as they were selected as the most probable candidates\cite{padovaniresconi14}. For example, if
$Y_{\nu\gamma} = 0.3$ then we expect $N_{\nu} \approx 3$ (4) for $2 < \en < 100$ PeV, which is well within the 2$\sigma$ limit for 0 events. 
\section*{Acknowledgments}
M.P. acknowledges support for this work by NASA through Einstein Postdoctoral 
Fellowship grant number PF3~140113 awarded by the Chandra X-ray 
Center, which is operated by the Smithsonian Astrophysical Observatory
for NASA under contract NAS8-03060.  E.R. is supported by a Heisenberg Professorship of the Deutsche
Forschungsgemeinschaft (DFG RE 2262/4-1).

\end{document}